\documentstyle{article}            %% varvi.   
\def\bc{\begin{center}}           \def\ec{\end{center}}
\def\beq{\begin{equation}}         \def\eeq{\end{equation}}
\def\bear{\begin{eqnarray}}       \def\eear{\end{eqnarray}}
\def\bt{\begin{tabular}}          \def\et{\end{tabular}}
\def\la{\langle}                  \def\ra{\rangle}
\def\dg{\dagger}                  \def\ci{\cite}
\def\lb{\label}                   \def\ld{\ldots}
                  \def\vs{\vspace}
\def\pr{\prime}                   
              \def\rar{\rightarrow}
        \def\lrar{\longrightarrow}
\def\td{\tilde}                   \def\pr{\prime}
\def\pd{\partial}                 
               \def\nn{\nonumber}
\def\ns{\normalsize}
\def\alf{\alpha}         
     \def\k{\kappa}
\def\Dlt{\Delta}         \def\eps{\epsilon}
\def\lam{\lambda}    \def\Lam{\Lambda}    \def\sig{\sigma}
\def\z{\zeta}      \def\vphi{\varphi}    
\def\ome{\omega}      \def\Ome{\Omega}    \def\tet{\theta}
      
\begin{document}

\title{ {\flushleft {\ns E-print quant-ph/9912084}\\
     {\ns Appeared in {\it Geometry, Integrability and Quantization,}}\\[-3mm] 
      {\ns Eds. I.M. Mladenov and G.L. Naber (Coral Press, Sofia 2000), 
	  p. 257-282} \\[-3mm]
      {\ns (Proc. Int. Conference, September 1-10, 1999, Varna)}\\[5mm]}
  \bf The Uncertainty Way of Generalization\\ of Coherent States}

\author{D.A. Trifonov\\
        Institute for Nuclear Research and Nuclear Energetics\\
        72 Tzarigradsko chauss\'ee, Sofia, Bulgaria }
\maketitle

\begin{abstract}
The three ways of generalization of canonical coherent states are briefly
reviewed and compared with the emphasis laid on the (minimum) uncertainty
way. The characteristic uncertainty relations, which include the
Schr\"odinger and Robertson inequalities, are extended  to the case of
several states. It is shown that the standard $SU(1,1)$ and $SU(2)$
coherent states are the unique states which minimize the second order
characteristic inequality for the three generators.  A set of states which
minimize the Schr\"odinger inequality for the Hermitian components of the
$su_q(1,1)$ ladder operator is also constructed.  It is noted that the
characteristic uncertainty relations can be written in the alternative
complementary form.
\end{abstract}

%section 1
\section{Introduction}

Coherent states (CS) introduced in 1963 in the pioneering works by Glauber
 and Klauder \cite{KlaSka} pervade nearly all branches of quantum physics
(see the reviews \ci{KlaSka}--\ci{AliAGM}). This important
overcomplete family of states $\{|\alf\ra\}$, $\alf \in {\mathbf C}$, can be
defined in three equivalent ways \cite{ZhaFenGil}: \\[-2mm]

(D1) As the set of eigenstates of boson destruction operator (the
ladder operator) $a$:\,\,
$a|\alf\ra = \alf|\alf\ra,$

(D2) As the orbit of the ground state $|0\ra$ ($a|0\ra = 0$) under the
action of the unitary displacement operators $D(\alf)=\exp(\alf a^\dg -
\alf^*a)$ (which realize ray representation of the Heisenberg--Weyl group
$H_1$) :\,\, $|\alf\ra = D(\alf)|0\ra$.

(D3) As the set of states which minimize the Heisenberg uncertainty relation
(UR) $(\Delta q)^2(\Delta p)^2 \geq 1/4$ for the Hermitian components
$q,\,p$ of $a$ ($a = (q+ip)/\sqrt{2}$) with equal uncertainties:\,\,
$(\Delta q)^2(\Delta p)^2 = 1/4,\,\, \Delta q = \Delta p$. Note that one
requires the minimization plus the equality of the two variances.

The overcompleteness property reads ($d^2\alf= d{\rm Re}\alf\,d{\rm Im}\alf$)
\beq\lb{1}     %eq.1
1 = \int|\alf\ra\la\alf|d\mu(\alf),\quad d\mu(\alf) = (1/\pi) d^2\alf.
\eeq
One says that the family $\{|\alf\ra\}$ resolves the unity operator with
respect to the measure $d\mu(\alf)$. The CS $|\alf\ra$ should be referred as
{\it canonical CS} \cite{KlaSka}.  The resolution unity property (\ref{1})
provides the important analytic representation (rep ), known as canonical CS
rep or Fock--Bargman analytic rep, in
which $a=d/d\alf,\,\, a^\dg = \alf$ and the state $|\Psi\ra$ is represented
by the function $\Psi(\alf) = \exp(|\alf|^2/2)\la\alf^*|\Psi\ra$.  In
1963-64 Klauder \cite{KlaSka} developed a general theory of the continuous
reps and suggested the possibility to construct overcomplete sets
of states using irreducible reps of Lie groups. Let us note that
the resolution unity property (\ref{1}) is not a defining one for the CS
$|\alf\ra$.

Correspondingly to the definitions (D1)--(D3) there are three
different ways (methods) of generalization of the canonical CS \ci{ZhaFenGil}:
The diagonalization of non-Hermitian operators (the {\it eigenstate way}, or
the ladder operator method \cite{Nie98}); The construction of Hilbert space
orbit by means of unitary operators ({\it orbit way} or the displacement
operator method \cite{Nie98}); The minimization of an appropriate UR
(the {\it uncertainty way}). The first two methods and especially
the second one (the orbit method) have enjoyed a considerable attention and
vast applications to various fields of physics \ci{KlaSka}--\ci{AliAGM},
while the third method is receiving a significant attention  only recently
-- see \ci{Trif94}--\ci{Trif98a}, \ci{FujFun}--\ci{Bjork} and references
therein. It is worth noting at the point that some authors were pessimistic
about the possibility of effective generalization of the third
defining property of canonical CS.

The aim of the present paper is to consider some of the new developments in
the third way (the uncertainty way) and their relationship to the first two
methods. We show that the Robertson \cite{SchRob} and other characteristic
inequalities \cite{TriDon} are those uncertainty relations which are
compatible with the generalizations of the ladder operator and displacement
operator methods to the case of many observables.

In section 2 we briefly review some of the main generalizations of the first
two defining properties of the canonical CS and the relationship between the
corresponding generalized CS.  Some emphasis is laid on the family of
squeezed states (SS) \ci{LouKni} and the Barut-Girardello CS (BG CS)
\cite{BG} and their analytic reps.  The canonical SS are the unique
generalization of CS for which the three definitions (D1), (D2), (D3) are
equivalently generalized.

Section 3 is devoted to the uncertainty way of generalization of CS. In
subsection 3.1 we consider the minimization of the Heisenberg and the
Schr\"odinger UR \cite{SchRob} for two observables and the relation of the
minimizing states to the corresponding group-related CS \cite{KlaSka}, on
the examples of $SU(2)$, $SU(1,1)$ and $SU_q(1,1)$. Here we note that the 
$SU(2)$ and  $SU(1,1)$ CS with lowest (highest) weight reference vector
minimize the Schr\"odinger inequality for the first two generators, while
the Heisenberg one is minimized in some subsets only. These group-related CS 
are particular cases of the corresponding minimizing states. A set of states
which minimize the Schr\"odinger inequality for the Hermitian
componenents of the $SU_q(1,1)$ ladder operator is also constructed.

In the subsection 3.2 the minimization of the Robertson \cite{Rob} and the
other {\it characteristic UR} \cite{TriDon} for several observables is
considered. 
In the case of the three generators (three observables) of $SU(1,1)$ (and
the $SU(2)$) we establish that the group-related CS with lowest (highest)
weight reference vector are the unique states which minimize the second and
the third characteristic UR for the three generators simultaneously. The
characteristic UR, in particular the known Robertson and the Schr\"odinger
ones, relate certain combinations of the second and first moment of the
observables in one and the same quantum state. Here we extend these
relations to the case of several states.  States which minimize the
characteristic UR are naturally called {\it characteristic uncertainty
states} (characteristic US
\footnote[1]{Let us list the abbreviations used in the paper: CS = coherent
state, SS = squeezed state, UR = uncertainty relation, US = uncertainty
state, BG = Barut-Girardello, and rep = representation.}).  The alternative
names could be (characteristic) intelligent states and (characteristic)
optimal US. The extended characteristic UR are also {\it invariant} under
the linear nondegenerate transformation of the observables as the
characteristic ones are.  It is shown that the characteristic UR can be
written in the {\it complementary form} \cite{Bjork} in terms of two positive
quantities less than the unity.  Finally it is noted that the positive
definite characteristic uncertainty functionals (for several observables)
can be used for the construction of distances between quantum states. In the
Appendix the proofs of the Robertson relation (after Robertson) and of the
uniqueness of the standard $SU(1,1)$ CS minimization of the second (and third)
order characteristic UR are provided. 

%%Section 2.
\section{The Eigenstate and Orbit Ways}

Canonical CS $|\alf\ra = D(\alf)|0\ra$ diagonalize the boson destruction
operator $a$, $[a,a^\dg]=1$. This was the first and seminal example of
diagonalizing of a non-Hermitian operator. We stress that the eigenstates of
$a$ and other non-Hermitian operators in this paper  are not orthogonal to
each other -- the term "diagonalization" is used for brevity and in analogy
to the case of diagonalization of Hermitian operators. The second example
was, to the best of our knowledge, the diagonalization of the complex
combination of boson lowering and raising operators $a,\,a^\dg$
($\alf\in {\mathbf C}$),
\ci{MMT}
\beq\lb{|alf;t>}     %eq.2
A(t)|\alf;t\ra = \alf|\alf;t\ra,\quad A(t) = u(t)a +
v(t)a^\dg = A(u,v).
\eeq
The operator $A(t)$ was constructed as a non-Hermitian invariant operator
for the quantum varying frequency oscillator with Hamiltonian  $H =
\left(p^2 + m^2\ome^2(t)q^2\right)/2m$,\,\, i.e. $A(t)$ had to obey the
equation $\pd A/\pd t - (i/\hbar)[A,H] = 0$ [$m$ is the mass, and $\ome(t)$
is the varying frequency; the case of varying mass $m(t)$ was
reduced to that of constant mass by the time transformation 
$t \rightarrow t^\pr = m\int^t d\tau/m(\tau)$]. 
For that purpose the parameter $\eps = (u-v)/\sqrt{\ome_0}$ was introduced 
and subjected to obey the classical oscillator equation
\beq\lb{eps}     %eq.3
\ddot{\eps} + \ome^2(t)\eps = 0.
\eeq
The boson commutation relation $[A,A^\dg]=1$ was  ensured by the Wronskian
$\eps^*\dot{\eps} - \eps\dot{\eps}^* =  2i$.  Then $\dot{\eps} = i(u+v)
\sqrt{\ome_0}$, $|u|^2 - |v|^2 = 1$, and the invariant takes the
form $A(t) = U(t)\left(u(0)a + v(0)a^\dg\right) U^\dg(t) \equiv
U(t)A(0)U^\dg(t),$ where $U(t)$ is the evolution operator, and the
eigenstates $|\alf;t\ra \equiv |\alf,u(t),v(t)\ra$ satisfy the Schr\"odinger
evolution equation.  One has
\beq\lb{|alf,u,v>}     %eq.4
|\alf,u(t),v(t)\ra = U(t)|\alf,u_0,v_0\ra,
\eeq
where $A(0)|\alf,u_0,v_0\ra = \alf|\alf,u_0,v_0\ra$ and $|u_0|^2 - |v_0|^2
= 1$.
This shows that the set $\{|\alf,u(t),v(t)\ra\}$ is an orbit through
$|\alf,u_0,v_0\ra$ of the evolution operator $U(t)$.

In the coordinate rep the wave functions $\la q|\alf,u(t),v(t)\ra$
take the form of an exponential of a quadratic \ci{MMT} ($m$ is the mass
parameter),
\bear\lb{<q|alf,u,v>}     %eq.5
\la q|\alf,u,v\ra =\frac{(m\ome_0/\pi\hbar)^{1/4}}
{(u-v)^{1/2}}\hspace{6.5cm}\nn \\
\times \exp\left[-\frac{m\ome_0}{2\hbar} \frac{v+u}{u-v}
\left(q - \left(\frac{2\hbar}{m\ome_0}\right)^{1/2}\frac{\alf}{u+v}\right)^2
- \frac 12\left(-\frac{u^* + v^*}{u+v}\alf^2 + |\alf|^2\right)\right].
\eear
These wave packets are normalized but not orthogonal to each other. 
They are
solutions to the wave equation for varying frequency oscillator if
$u=(\eps\sqrt{\ome_0}-i\dot{\eps}/\sqrt{\ome_0})/2$,
$v = -(\eps\sqrt{\ome_0}+i\dot{\eps}/\sqrt{\ome_0})/2$,
and $\eps$ is any solution of (\ref{eps}).
Note that the time dependence is embedded completely in $u$ and $v$ (or,
equivalently, in $\eps$ and $\dot{\eps}$) which justifies the notation
$|\alf;t\ra = |\alf,u,v)\ra$.  For other systems the invariant
$A(t)=U(t)A(0)U^\dg(t)$ is not linear in $a$ and $a^\dg$ and its eigenstates
are no more of the form $|\alf,u,v\ra$ \ci{Trif93}.  Therefore the term
"coherent states for the nonstationary oscillator" for $|\alf;t\ra =
|\alf,u,v\ra$ \ci{MMT} is indeed adequate. Time evolution of an initial
$|\alf,u_0,v_0\ra$ for general quadratic Hamiltonian system was studied in
greater detail in \ci{Yuen}, where eigenstates of $ua + va^\dg$ were denoted
as $|\alf\ra_g$. The invariant $A(t)$ in \ci{MMT} coincides with the boson
operator $b(t)$ in \ci{Yuen}.

The states (\ref{<q|alf,u,v>}) represent the time evolution of the canonical
CS $|\alf\ra$ if the initial conditions \ci{MMT} $\eps(0) =
1/\sqrt{\ome_0},\,\, \dot{\eps}(0) = i\sqrt{\ome_0}$ are imposed (then
$u(0)=1,\, v(0)=0$). Under these conditions
$|\alf,u(t),v(t)\ra = U(t)|\alf\ra$,
i.e. the set of $|\alf,u(t),v(t)\ra$ becomes an $SU(1,1)$
orbit through the initial CS $|\alf\ra$, since the Hamiltonian of the varying
frequency oscillator is an element of the $su(1,1)$ algebra
in the rep  with Bargman index $k=1/4,3/4$.  The $SU(1,1)$
generators $K_i$ in this rep read\,\, ($K_\pm = K_1\pm iK_2$)
\beq\lb{1moderep}     %eq.6
K_3 = \frac 12 a^\dg a + \frac 14,\quad K_- = \frac 12a^2,\quad 
K_+ = \frac 12 a^{\dg 2}.
\eeq
The parameters $u,\, v$ are in a direct link to the $SU(1,1)$ group
parameters, and $\alf$ -- to the Heisenberg--Weyl group. The whole family of
$|\alf,u,v\ra$, can be considered as an orbit through the ground state
$|0\ra$ of the unitary operators of the semidirect product $SU(1,1)\wedge
H_1$ \ci{Trif93}. Thus the two definitions (D1) and (D2) here are
equivalently generalized. It has been shown \ci{Trif93} that the third
definition is also equivalently generalized on the basis of the Schr\"odinger
UR (see next section).

The set $\{|\alf,u,v\ra,\,\,u,v - {\rm fixed}\}$ resolves the unity operator
with respect to the same measure as in the case (\ref{1}) of canonical CS
\ci{MMT}: $1 = (1/\pi)\int d^2\alf\,|\alf,u,v\ra\la v,u,\alf|$.

A second family of orthonormalized states $|n;t\ra = |n,u,v\ra$ was
constructed in \cite{MMT} as eigenstates of the quadratic in $a$ and $a^\dg$
Hermitian invariant $A^\dg(t) A(t) = (ua + va^\dg)^\dg (ua + va^\dg)$ which
is an element of the Lie algebra $su(1,1)$. Note that any power of $A$ and
$A^\dg$ is also an invariant. $A^\dg A$ coincides with the known
Ermakov--Lewis invariant.  For the $N$-dimensional quadratic system there
are $N$ linear in $a_\mu$ and $a^\dg_\mu$ invariants $A_\mu(t) =
u_{\mu\nu}a_\nu + v_{\mu\nu}a^\dg_\nu \equiv A_\mu(u,v)$ ($\mu,\nu =
1,2,\ldots N$), which were simultaneously diagonalized \ci{HMMT},
\beq\lb{2a}     %eq.7
A_\mu(u,v)|\vec{\alf},u,v\ra = \alf_\mu|\vec{\alf},u,v\ra,
\eeq
In different notations exact solutions to the Schr\"odinger equation for the
nonstationary oscillator have been previously obtained e.g. by Husimi
\ci{HCh} and for nonstationary general $N$-dimensional Hamiltonian by
Chernikov \ci{HCh}, but with no reference to the eigenvalue problem of
the invariants $ua+va^\dg$ and/or $(ua+va^\dg)^\dg(ua+va^\dg)$.  Eigenstates
of other quadratic in $a$ and $a^\dg$ operators were later considered in
many papers, the general one-mode quadratic form being diagonalized by Brif
(see \ci{Brif96} and references therein).

By means of the known BCH formula for the transformation $S(\z)aS^\dg(\z)$
with
$S(\z) = \exp[\z K_+ - \z^*K_-],\quad K_- =a^2/2,\,\, K_+=  a^{\dg 2}/2$,
the solutions $|\alf,u,v\ra$ are immediately brought, up to a phase factor,
to the form of famous Stoler states $|\alf,\z\ra = S(\z)|\alf\ra$ \ci{Stol}:
\beq\lb{Stolform}     %eq.8
|\alf,u,v\ra = e^{i{\rm arg}\,u}\,\exp(\z K_+ - \z^*K_-) |\alf\ra,
\eeq
where $|\z| ={\rm arcosh}|u|$ and ${\rm arg}\,\z ={\rm arg}\,v-{\rm arg}\,u$.
Yuen \ci{Yuen} called the eigenstates $|\alf,u,v\ra$ of $ua+va^\dg$  {\it
two photon CS} and suggested that the output radiation of an ideal
monochromatic two photon laser is in a state $|\alf,u,v\ra$. In
\ci{Hollen} these states were named {\it squeezed states} (SS) to reflect
the property of these states to exhibit fluctuations in $q$ or $p$ less
than those in CS $|\alf\ra$. They were intensively studied in quantum
optics and are experimentally realized (see refs in \ci{LouKni,ZhaFenGil}).
The eigenstates $|n,u,v\ra$ of $(ua+va^\dg)^\dg(ua+va^\dg)$ became known as
 squeezed Fock states ($|n\!=\!0,u,v\ra$ -- squeezed vacuum) and the
operator $S(\z)$ -- (canonical) {\it squeeze operator} \ci{LouKni,ZhaFenGil}.
Eigenstates $|\vec{\alf},u,v\ra$, eq.  (\ref{2a}), became known as multimode
(canonical) SS.

Noting that the variance $(\Delta X)^2$ of a Hermitian operator $X$ in a
state $|\Psi\ra$ equals zero iff $|\Psi\ra$ is an eigenstate of $X$ so it
was suggested \ci{Trif94} to construct SS for {\it arbitrary two
observables} $X_1$ and $X_2$, in analogy to the canonical SS
$|\alf,u,v\ra$, as eigenstates of their complex combination $\lam X_1
+iX_2$, $\lambda \in {\mathbf C}$ (or equivalently $uA + vA^\dg$,
$A=(X_1-iX_2)$), since if in such eigenstates $\lambda \rar 0$ ($\lambda
\rar \infty $) then $\Delta X_2 \rar 0$ ($\Delta X_1 \rar 0$) \ci{Trif94}.

Radcliffe \ci{Radclif} and Arecchi et al \ci{Gilmore} introduced and
studied the $SU(2)$ analog $|\tet,\vphi;j\ra$ of the states
$|\alf\!=\!0,u,v\ra$ in the similar form to that of Stoler states
(\ref{Stolform}) (the displacement operator form) ($J_\pm = J_1\pm iJ_2$),
\beq\lb{SU2CS}     %eq.9
|\tet,\vphi\ra = \exp(\z J_+ -\z^*J_-)|j,-j\ra =
\left(\frac{-1}{1+|\tau|^2}\right)^j\, e^{\tau J_+}|j,-j\ra 
\equiv |\tau;j\ra,
\eeq
where  $|j,m\ra$ ($m=-j,-j+1,\ldots,j$, $j=1/2,1,\ldots$) are the standard
Wigner--Dicke states, the operators $J_1$,
$J_2$ and $J_3$ are the Hermitian generators of $SU(2)$,
$\tau = \exp(-i\vphi){\rm tan}(\tet/2)$, $\z = (\tet/2)\exp(-i\vphi)$ and
$\vphi$ and $\tet$ are the two angles in the spherical coordinate
system. The system $\{|\tet,\vphi\ra\}$ is overcomplete \ci{Gilmore},
\beq\lb{9}     %eq.10
1 = [(2j+1)/4\pi]\int d\Ome |\tet,\vphi\ra\la\vphi,\tet|,
\eeq
where $d\Ome = \sin\tet d\tet d\vphi$.
The states $|\tet,\vphi\ra \equiv |\tau;j\ra$ are known as {\it spin
CS} \ci{Radclif} or {\it atomic CS} (Bloch states) \ci{Gilmore}.

The results of \ci{Radclif,Gilmore} about the $SU(2)$ CS have been extended
to the noncompact group $SU(1,1)$ and to any Lie group $G$ as well by
Perelomov \ci{Perel}, who succeeded to prove the Klauder suggestion for
construction of overcomplete families of states using unitary irreducible
reps of a Lie group $G$. If $T(g)$ is an irreducible unitary
rep of $G$, $|\Psi_0\ra$ is a fixed vector in the rep
space, $H$ is stationary subgroup of $|\Psi_0\ra$ (that is $T(h)|\Psi_0\ra =
\exp[i\alf(h)]|\Psi_0\ra$) then the family of states $|x\ra = T(s(x))
|\Psi_0\ra$, where $s(x)$ is a cross section in the group fiber bundle,
$x\in {\cal X} = G/H$, is overcomplete, resolving the unity with respect
to the $G$-invariant measure on ${\cal X}$,
\beq\lb{ru}     %eq.11
1 = \int |x\ra\la x|d\mu(x),\quad d\mu(g\cdot x) = d\mu(x).
\eeq
Such states were called  generalized CS and denoted as CS of the
type $\{T(g),\Psi_0)$\} \ci{Perel}. It is worth noting that an other type of
"generalized CS" was previously introduced by Titulaer and Glauber
(see the ref. in \ci{KlaSka}) as the most general states which satisfy the
Glauber field coherence condition.  Therefore we adopt the notion
"group-related CS" for the generalized CS of the type $\{T(g),\Psi_0\}$
\ci{KlaSka}.  The Perelomov $SU(1,1)$ CS $|\z;k\ra$ for the discrete series
$D^+(k)$ with the reference vector $|\Psi_0\ra = |k,k\ra$\,\, ($K_-|k,k\ra =
0$, $K_3|k,k\ra = k|k,k\ra$) have quite similar form to that of spin CS
(\ref{SU2CS}) and Stoler states (\ref{Stolform}),
\beq\lb{SU11CS}     %eq.12
|\z,k\ra = \exp(\z K_+ - \z^* K_-)|k,k\ra = (1-|\xi|^2)^k\,e^{\xi
K_+}|k,k\ra \equiv |\xi;k\ra,
\eeq
where $|\xi|= {\rm tanh}|\z|$, arg$\xi = -{\rm arg}\z+\pi$. The $SU(1,1)$
and $SU(2)$ invariant resolution unity measures for these sets of
states are  ($k\geq 1/2$) \ci{Perel}
\beq\lb{measures}     %eq.13
d\mu(\xi) = [(2k-1)/\pi]d^2\xi/(1-|\xi|^2)^2,\qquad
d\mu(\tau) = [(2j+1)/\pi]d^2\tau/(1+|\tau|^2)^2.
\eeq
The $SU(1,1)$ reps with $k=1/2$ and $k=1/4$ are not square
integrable against the invariant measure $d\mu(\xi)$. The whole family of
canonical SS $|\alf,u,v\ra$, eqs. (\ref{|alf;t>}), (\ref{|alf,u,v>}), 
remains stable (up to a phase factor) 
under the action of unitary operators of the semidirect
product $SU(1,1)\wedge H_1$.  However it does not resolve the identity
operator with respect to the corresponding $SU(1,1)\wedge H_1$ invariant
measure \ci{Trif93}.  Noninvariant resolution unity measures for the set of
canonical SS were found in \ci{Trif93,Beckers}.
The overcompleteness property of the CS $|\tau;j\ra$ and
$|\xi;k\ra$ provide the analytic reps in the complex plain and in
the unit disk respectively which were successfully used by Brif \ci{Brif97})
for diagonalization of the general complex combinations of the $SU(2)$ and
$SU(1,1)$ generators. The $SU(1,1)$ analytic rep in the unit disk
was also considered in \ci{Vourd90,BrifBen}.

A lot of attention is paid in the physical literature, especially in
quantum optics, to the group-related CS for $SU(2)$ and $SU(1,1)$ in their
one- and two-mode boson reps, such as the Schwinger two mode reps (see
\ci{KlaSka,LouKni,ZhaFenGil, LuiPer,Vourd90} and references therein), and
the one-mode Holstein--Primakoff reps (see e.g. \ci{Vourd90,VouBrif} and
references therein).

An extension of the group-related CS, compatible with the resolution of the
identity, can be obtained if the stationary subgroup $H\subset G$ in
Gilmore--Perelomov scheme is replaced by other closed subgroup (references
[1]-[8] in \ci{AliAGM}).  Significant progress is achieved recently
\ci{AliAGM} in the construction of more general type of continuous families
of states (called also CS \ci{AliAGM}) which satisfy the generalized
overcompleteness relation $B = \int |x\ra\la x|d\mu(x)$, where $B$ is a
bounded, positive and invertible operator. When $B=1$ the Klauder definition
of general CS (overcomplete family of states) \ci{KlaSka} is recovered. \\

Along the line of generalization of the eigenvalue property (D1) of the
canonical CS the next step was made in 1971 by Barut and Girardello in
\ci{BG}, where the Weyl lowering generator $K_-$ of $SU(1,1)$ in the
discrete series $D^\pm(k)$ was diagonalized explicitly,
\beq\lb{BGCS}     %eq.14
K_-|z;k\ra = z|z;k\ra,\quad  |z;k\ra = N_{BG}\sum_{n = 0}^{\infty}
\frac{z^{n}}{\sqrt{n!\Gamma(2k+n)}} |k,k+n\rangle.
\eeq
The family $\{|z;k\ra\}$ resolves the unity operator,
$1 = \int|z;k\ra\la k,z|d\mu(z,k)$, the resolution unity measure being
\beq\lb{BGmeasure}     %eq.15
 d\mu(z,k) = \frac{2}{\pi}
\left(N_{BG}\right)^{-2} |z|^{2k-1} K_{2k-1}(2|z|)\, d^{2} z,
\eeq
where $K_\nu(x)$ is the modified Bessel function of the third kind 
\ci{Stegun}.  The identity operator resolution (\ref{BGmeasure}) provides a
new analytic rep in Hilbert space \ci{BG}.  The measure $d\mu(z,k)$, eq.
(\ref{BGmeasure}), is {\it not invariant} under the action of the $SU(1,1)$
on ${\mathbf C} \ni z$.  In the Barut--Girardello (BG) rep  states $|\Psi\ra$
are represented by functions $F_{BG}(z) = \la k,z^{\ast}|\Psi\ra/ 
N_{BG}(|z|,k)$ which are of the growth $(1,1)$.  The
orthonormalized states $|k,k+n\ra$ are represented by monomials
$z^n/\sqrt{n!(2k)_n}$, $(2k)_n = \Gamma(2k+n)/\Gamma(2k)$.  The $SU(1,1)$
generators $K_\pm$ and $K_3$ act in the space ${\cal H}_k$ of analytic
functions $F_{BG}(z)$ as linear differential operators
\beq\lb{BGrep}     %eq.16
K_{+} =  z,  \quad K_{-} = 2k \frac{d}{dz} + z \frac{d^{2}}{dz^{2}},
 \quad K_{3} = k + z \frac{d}{dz}.
\eeq
Originally established for the discrete series $D^+(k)$, $k=1/2,1,\ldots$ the
BG rep is in fact valid for any positive index $k$.
Recently this rep has been used to diagonalize the complex
combination $uK_- + vK_+$ of the Weyl operators $K_\pm$ \ci{Trif94} and the
general element of $su(1,1)$ as well \ci{Trif96,Trif97,Brif96,Brif97}.  The
relations between BG rep and the Fock-Bargmann analytic rep (also
called canonical CS rep) have been established in \ci{BVM} (the
case of $k=1/4,\,3/4$) and \ci{Trif98a} (the cases of $k = 1/2,1,3/2,
\ldots$).
The BG-type analytic rep was recently extended to the
algebras $u(N,1)$ \ci{FujFun} and $u(p,q)$ in their boson realizations
\ci{Trif98a}.  The BG-type CS for these and any other (noncompact) semisimple
Lie algebra are defined \ci{Trif98a} as common eigenstates of the mutually
commuting Weyl ladder operators.

The BG CS $|z;k\ra$ can be also defined according to the third definition
(D3) on the basis of the Heisenberg relation for $K_1$ and $K_2$. For this
family the generalization of the definition (D2) does not exist \ci{Trif98b}.

The ladder operator method was extended to the deformed quantum oscillator
in \ci{Bied}, where the $q$-deformed boson annihilation operator $a_q$,
\beq\lb{a_q}     %eq.17
a_q a^\dg_q - q a^\dg_q a_q = q^{-\hat{n}},\quad [\hat{n},a^\dg_q]=a^\dg_q,
\quad q > 0,
\eeq
has been diagonalized, the eigenstates $|\alf\ra_q$  being called
"$q$-CS" or CS for the quantum Heisenberg--Weyl group $h_q(1)$,
\beq\lb{|alf>_q}     %eq.18
|\alf\ra_q = {\cal N}\exp_q(\alf a_q^\dg)|0\ra =
{\cal N}\sum_n^\infty\frac{\alf^n}{\sqrt{[n]_q!}}|n\ra,\quad
{\cal N}=\exp_q(-|\alf|^2),
\eeq
where $\exp_q(x) = \sum x^n/[n]_q!$,\,\,\, $[n]_q! = [1]_q\ldots[n]_q$, 
$a^\dg a|n\ra = n|n\ra$ (and $a^\dg_q a_q|n\ra = [n]_q|n\ra$). 
The "classical limit" is obtained at $q=1$: $a_{q=1} = a$.
The $q$-SS have been constructed in the first paper of \ci{Solom} as states
$|v\ra_q$ annihilated by the linear combination $a_q + va^\dg_q$,  in
analogy to the case of canonical squeezed vacuum states $|\alf\,=\,0,u,v\ra$:
$(a_q + va^\dg_q)|v\ra_q = 0$. It was noted \ci{Solom} that both $q$-CS and
$|v\ra_q$ can exhibit squeezing in the quadratures of the (ordinary) boson
operator $a$. Group-related type CS associated with the $q$-deformed
algebras $su_q(2)$, $[J_-(q),J_+(q)] = - [2J_3]_q,\quad [J_3,J_\pm(q)] = \pm
J_\pm(q),\lb{su2_q}$, and $su_q(1,1)$, $[K_-(q),K_+(q)] = [2K_3]_q,\quad
[K_3,K_\pm(q)] = \pm K_\pm(q)$, in their Holstein--Primakoff realizations in
terms of $a_q$,

\begin{eqnarray}     %eq.19  %eq.20
J_-(q) = a_q\sqrt{[-\hat{n}+ 2\k+1]_q},\quad  J_+(q) =
\sqrt{[-\hat{n}+2\k+1]_q}\, a^\dg_q,\quad  J_3 = \hat{n} -
\k,\lb{su2_qHP}\\[2mm]
K_-(q) = a_q\sqrt{[\hat{n}+ 2\k - 1]_q},\quad  K_+(q) =
\sqrt{[\hat{n}+2\k-1]_q}\, a^\dg_q,\quad  K_3 = \hat{n}+\k,\lb{su11_qHP}
\end{eqnarray}
were constructed and discussed in \ci{Kulish,Solom} ($\k=1/2$ in \ci{Kulish}
and any $\k$ in \ci{Solom}). Here $[x]_q \equiv (q^x-q^{-x})/(q-q^{-1})$.
These $su(2)$ and $su(1,1)$ $q$-CS are defined similarly to the ordinary
group-related CS (\ref{SU2CS}) and (\ref{SU11CS}) with $J_i$, $K_i$, $n!$ and
$(x)_n$ replaced by their $q$-generalizations \ci{Kulish,Solom}. Their
overcompleteness relations (in terms of the Jackson $q$-integral) can be
found in \ci{Ellin}, the corresponding resolution unity measures being the
$q$-deformed versions of $d^2\alf$ and (\ref{measures}): $d\mu_q(\alf) =
d^2_q\alf/\pi$,
\beq\lb{q-measures}     %eq.21
d\mu_q(\tau)=\frac{[2j+1]_q}{ _q\la j;\tau|\!|\tau;j\ra_q^2}d^2_q\tau,\quad
d\mu_q(\xi) = \frac{[2k-1]_q}{_q\la k;\xi|\!|\xi;k\ra_q^{-2}}d^2_q\xi,
\eeq
where $|\!|\tau;j\ra_q = \exp_q(\tau J_+(q))|j,-j\ra$, 
 $|\!|\xi;j\ra_q = \exp_q(\xi K_+(q))|k,k\ra$. 
The Barut-Girardello $q$-CS (eigenstates of $K_-(q)$) are constructed in the
first paper of \ci{Kulish}. The ladder operator formalism for several kinds
of one- and two-mode boson states is considered recently in \ci{Wang}.  For
further development in the field of $q$-deformed CS see e.g. \ci{Ellin, 
Solom3Oh}. For CS related to supergroups ({\it super-CS}) see e.g. \ci{Fatyga}.
The canonical SS can be regarded as super-CS related to the orthosymplectic
supergroup $OSp(1/2,R)$ \ci{Trif92}.

%%Section 3
\section{The Uncertainty Way}

 %subsection 3.1
 \subsection{The Heisenberg and the Schr\"odinger UR}

Canonical CS $|\alf\ra$ (and only they) minimize the Heisenberg uncertainty
relation with equal uncertainty of the two (dimensionless) canonical
observables $p$ and $q$: in $|\alf\ra$ the two variances are equal and
$\alf$- independent, $(\Dlt p)^2 = 1/2 = (\Dlt q)^2$. $1/2$ is the lowest
level at which the equality $(\Dlt p)^2 = (\Dlt q)^2$ can be maintained.
Therefore the set of $|\alf\ra$ is {\it the set of $p$-$q$ minimum
uncertainty states}. The CS related to any other two (or more) {\it
noncanonical observables} $X_1$ and $X_2$ {\it are not} with minimal and
equal uncertainties -- the lowest level of the equality $(\Dlt X_1)^2 =
(\Dlt X_2)^2$ can be reached on some subsets only. For example, in the
$SU(1,1)$ CS $|\xi;k\ra$ the variances of the generators $K_1$ and $K_2$ for
$\xi\neq 0$ are always greater than their value in the lowest weight vector
state $|k,k\ra$: $\Dlt K_{1,2} (\xi) > \Dlt K_{1,2} (0) = \sqrt{k/2}$
\ci{Trif94}. The Heisenberg inequality for $K_1$ and $K_2$ is minimized in
the subsets of states with Re$\xi=0$ and/or Im$\xi=0$ only, but the
uncertainties $\Dlt K_{1}(\xi)$ and $\Dlt K_2(\xi)$ (calculated in
\ci{NikTrif}) are never equal unless $\xi = 0$.  Similar is the uncertainty
status of the spin CS ($SU(2)$ related CS) $|\tau;j\ra$.

It turned out \ci{Trif94} that the above $SU(1,1)$ and $SU(2)$ group related
CS minimize, for any values of the parameters $\xi$ and $\tau$, the more
precise uncertainty inequality of Schr\"odinger (called also
Schr\"odinger--Robertson inequality) \ci{SchRob},
\beq\lb{SUR}     %eq.22
(\Dlt X_1)^2 (\Dlt X_2)^2 \geq \frac 14\left|\la[X_1,X_2]\ra\right|^2 +
(\Dlt X_1X_2)^2,
\eeq
where $\la X\ra$ is the mean value of $X$, and $\Dlt X_1X_2 \equiv \la
X_1X_2+X_2X_1\ra/2 -\la X_1\ra \la X_2\ra$ is the covariance of $X_1$ and
$X_2$. However the sets of states which minimize (\ref{SUR}) for $K_{1,2}$
and $J_{1,2}$ are much larger than the sets of the corresponding
group-related CS $|\xi;k\ra$ and $|\tau;j\ra$ -- these larger sets have
been constructed in \ci{Trif94} as eigenstates of the general {\it
complex} combinations of the ladder operators $K_\pm$ and $J_\pm$
correspondingly since the necessary and sufficient condition for a state
$|\Psi\ra$ to minimize (\ref{SUR}) was realized to be the eigenvalue
equation
\beq\lb{OUS}     %eq.23
[u(X_1-iX_2) + v (X_1 + iX_2)]\,|\Psi\ra = z|\Psi\ra.
\eeq
The minimizing states should be denoted by $|z,u,v;X_1,X_2\ra$ and called
Schr\"odinger {\it $X_1$-$X_2$ optimal uncertainty states} (optimal US). The
other names already used in the literature are {\it generalized (or
Schr\"odinger) intelligent states} \ci{Trif94,Trif96}, {\it correlated CS}
\ci{DKM} and {\it Schr\"odinger minimum uncertainty states} \ci{Trif93}. The
minimization of the inequality (\ref{SUR}) for canonical $p$ and $q$ was
considered in detail in \ci{DKM}, where the minimizing states were called
{\it correlated CS}. The latter coincides with the canonical SS
$|\alf,u,v\ra$ \ci{Trif93}.  In the optimal US the uncertainties $\Dlt X_1$,
$\Dlt X_2$ are minimal in the case of $X_1=p$, $X_2=q$ only. Therefore the
frequently used term "minimum uncertainty states"
\ci{Trif93,Brif96,Trif96,HilNag,Nie98,LuiPer}
is generally not in its direct meaning.  The term {\it intelligent states}
was introduced in \ci{ArChalSal} on the example of Heisenberg inequality for
$J_{1,2}$.  States $|\Psi\ra$ for which the product functional
$U[\Psi]\equiv (\Dlt X_1)^2(\Dlt X_2)^2$ is stationary under arbitrary
variation of $|\Psi\ra$ \ci{Jackiw} were called by Jackiw {\it critical}.
Obviously there is no commonly accepted name for the states which minimize
an uncertainty inequality  -- the "optimal uncertainty states" is one more
attempt in searching for more adequate name.

In the solutions $|z,u,v;X_1,X_2\ra$ to (\ref{OUS}) the three second moments
of $X_1$ and $X_2$ are expressed in terms of the mean of their commutator
\ci{Trif94} (note that in \ci{Trif94} $\lambda,\,z^\pr$ parameters were used
instead of $u,v,z$: $\lambda = (v+u)/(v-u),\,\,z^\pr = z/(v-u)$),
\beq\lb{OUSmeans}     %eq.24
\left.\begin{tabular}{ll}
$\displaystyle
(\Dlt X_1)^2 = \frac{|u-v|^2}{|u|^2-|v|^2}\, C_{12},\quad
(\Dlt X_2)^2 = \frac{|u+v|^2}{|u|^2-|v|^2}\, C_{12},$  & \\[5mm]
$\displaystyle
\Dlt X_1X_2 = \frac{2{\rm Im}(u^*v)}{|u|^2-|v|^2}\,C_{12},\qquad C_{12}
= \frac i2\la[X_1,X_2]\ra.$   &
\end{tabular} \right\}
\eeq
These moments  satisfy the equality in (\ref{SUR}) identically with respect
to $z,u,v$.  From $(\Dlt X)^2 \geq 0$ and (\ref{OUSmeans}) it follows that if
the commutator $i[X_1,X_2]$ is positive (negative) definite then normalized
eigenstates of $u(X_1-iX_2) + v (X_1 + iX_2)$ exist for $|u|>|v|$
($|u|<|v|$) only \ci{Trif94}. In such cases one can rescale the parameters
and put $|u|^2-|v|^2 =1$ ($|u|^2-|v|^2 =-1$) as one normally does in the
canonical case of $X_1=p,\,X_2=q$.

In order to establish the connection of $K_1$-$K_2$ and $J_1$-$J_2$
optimal US $|z,u,v;K_1,K_2\ra \equiv |z,u,v;k\ra$ and $|z,u,v;J_1,J_2\ra
\equiv |z,u,v;j\ra$ with the displacement operator method consider the
operators
\bear\lb{displaceoper}     %eq.25   %eq.26
K^\pr_3 = \frac{i}{2}\sqrt{uv}\left(uK_- + vK_+\right),\quad
K^\pr_{\pm} = iK_3 \mp \left(\sqrt{u/v}\,K_-
-\sqrt{v/u}\,K_+\right),\\
J^\pr_3 = \frac{1}{2}\sqrt{uv}\left(uJ_- + vJ_+\right),\quad
J^\pr_{\pm} = J_3 \mp \left(\sqrt{u/v}\,J_- -\sqrt{v/u}\,J_+ \right),
\eear
which realize non-Hermitian reps of the algebras $su(1,1)$ and $su(2)$ with
the same indices $k$ and $j$. Therefore $(K^\pr_\pm)^n$ \,$\left((J^\pr_\pm)
^n\right)$ displace the eigenvalue $z$ of $uK_- + vK_+$\, ($uJ_- + vJ_+$) by
$\pm n$. If one could properly define noninteger powers of $K^\pr_\pm$\,
($J^\pr_\pm$) (to be considered elsewhere) one might write $|z,u,v;k\ra =
{\cal N}_1(K^\pr_\pm)^z|0,u,v;k\ra$\, ($|z,u,v;j\ra = 
{\cal N}_2(J^\pr_\pm)^z|0,u,v;j\ra$), where ${\cal N}_{1,2}$ are
normalization constants. 
In slightly different notations the operators $J^\pr_3,\,J^\pr_\pm$ were
introduced by Rashid \ci{Rashid}. 

An important physical property of the states $|z,u,v;X_1,X_2\ra$
is that they can exhibit arbitrary strong squeezing of the
variances of $X_1$ and $X_2$ when the parameter $v$ tend to $\pm u$, i.e.
$\Dlt X_{1,2} \lrar 0$ when $v \lrar \pm u$ \ci{Trif94}. Therefore the
families of $|z,u,v;X_1,X_2\ra$  are the {\it $X_1$-$X_2$ ideal SS}.
The canonical SS $|\alf,u,v\ra$ are $p$-$q$ ideal SS, while the
group-related CS $|\tau;j\ra$ and $|\xi;k\ra$ are not. 
Explicitly the families of $|z,u,v;X_1,X_2\ra$ are constructed for the
generators $K_i$-$K_j$ and $J_i$-$J_j$ of $SU(1,1)$
\ci{Trif94,Trif96,Brif97} and $SU(2)$ \ci{ArChalSal, Rashid,Brif97} (in
\ci{ArChalSal,Rashid} with no reference to the inequality (\ref{SUR})). It
is worth noting an important application of the $K_i$-$K_j$ and $J_i$-$J_j$
optimal US (intelligent states) in the quantum interferometry: the $SU(1,1)$
and $SU(2)$ optimal US which are not group-related CS can greatly improve
the sensitivity of the $SU(2)$ and $SU(1,1)$ interferometers as shown by
Brif and Mann \ci{LuiPer}.  Schemes for generation of $SU(1,1)$ and $SU(2)$
optimal US of radiation field can be found e.g. in \ci{Trif98b,LuiPer}.

Schr\"odinger optimal US can be constructed also for the two Hermitian
quadratures $K_1(q),\,K_2(q)$ ($J_1(q),\,J_2(q)$) of the ladder
operators of $q$-deformed $su_q(1,1)$ ($su_q(2)$). Let us consider here
the case of $su_q(1,1)$. The $K_1(q)$-$K_2(q)$ optimal US $|z,u,v;k\ra_q$
have to obey (\ref{OUS}) with $X_1=K_1(q)$ and $X_2=K_2(q)$.  We put
\beq\lb{|zuv;k>q}     %eq.27
|z,u,v;k\ra_q = {\cal N}_q|\!|z,u,v;k\ra_q = {\cal N}_q\sum_n g_n(z,u,v,q,k)
|k,k+n\ra,
\eeq
and substitute this in (\ref{OUS}). Using the actions
$K_-(q)|k,k+n\ra = \sqrt{[n][2k+n-1]}|k,k+n-1\ra$, and
$K_+(q)|k,k+n\ra = \sqrt{[n+1][2k+n]}|k,k+n-1\ra$
we get the recurrence relations for $g_n$,
\beq\lb{q_recurrence}     %eq.28
u\sqrt{[n+1][2k+n]}\,g_{n+1} + v\sqrt{[n+1][2k+n]}\,g_{n-1} = zg_n.
\eeq
The solution $g_n(z,v,u,q,k)$ to these recurrence relations is a polynomial
in $z/u$ and $v/u$,
\beq\lb{g_n}     %eq.29
g_n(z,u,v,q,k) =  \sum_{m=0}^{{\rm int}(n/2)}
p_{n,m}(k,q)\left(\frac{z}{u}\right)^{n-2m}\left(-\frac{v}{u}\right)
^{m},
\eeq
where int$(n/2)$ is the integer part of $n/2$. The particular case of $v=0$
was solved in \ci{Kulish}, $g_n(z,q,k) =  z^n/\sqrt{[n]!([2k])_n}$.  Here we
wright down the solution for the subset of $z=0$,
\beq\lb{z=0}     %eq.30
g_{2n+1}(u,v,q) = 0,\quad g_{2n}(u,v,q) = \left(-\frac{v}{u}\right)^n
\left(\frac{[2n-1]!!\,(\!([2k])\!)_{2n}}{[2n]!!\,(\!([2k+1])\!)_{2n}}\right)^
{\frac 12},
\eeq
and for $q=1$,
\beq\lb{q=1}     %eq.31
g_n(z,u,v,k) = \left(-\frac{l(u,v)}{2u}\right)^n \sqrt{\frac{(2k)_n}{n!}}\,
_2F_1\left(k+\frac{z}{l(u,v)},-n;2k;2\right),
\eeq
where $l(u,v) = 2\sqrt{-uv}$, $(\!([x])\!)_{2n} = [x][x+2]\ldots[x+2n-2]$
and $_2F_1(a,b;c;z)$ is the Gauss hypergeometric function. The normalization
condition is $|v| < |u|$.  The BG CS are recovered at $v=0,\,u=1$. The
construction of $g_n(z,u,v,q,k)$ in the general case is postponed until the
next publication.
%%\vs{5mm}

%subsection 3.2
\subsection{ The Robertson Inequality and the Characteristic UR}

Compared to the Heisenberg uncertainty relation the Schr\"odinger one, eq.
(\ref{SUR}), has the important advantage to be {\it invariant} under
nondegenerate linear transformations of the  two observables involved.
Indeed the relation (\ref{SUR}) can be rewritten  in the following invariant
form \ci{Rob} ${\rm det}\,\sigma(\vec{X}) \geq {\rm det}\, C(\vec{X})$,
where $\vec{X}$ is the column of $X_1$ and $X_2$, $\vec{X} = (X_1,X_2)$, and
\beq\lb{sigma,C}     %eq.32
C(\vec{X}) =
-\frac{i}{2}\left(\matrix{\,\,\,0\quad \quad \la[X_1,X_2]\ra\\[2mm]\cr
\la[X_2,X_1]\ra\quad \,\,\,0\quad}\right),
\qquad
\sigma(\vec{X}) = \left(
\matrix{\Dlt\,\!X_1X_1 \quad\Dlt\,\!X_1X_2\\[2mm]\cr
\Dlt\,\!X_2X_1\quad \Dlt\,\!X_2X_2}\right).
\eeq
$\sigma(\vec{X})$ is called the uncertainty (the dispersion) matrix for
$X_1$ and $X_2$. In order to symmetrize notations we have denoted in
(\ref{sigma,C}) the variance $(\Dlt X_i)^2$ as $\Dlt X_iX_j$. So
$\sigma_{ij} = \Dlt X_i X_j$ and $C_{kj} = -(i/2)\la[X_k,X_j]\ra$.  Under
linear transformations $\vec{X} \longrightarrow \vec{X}^\pr = \Lam \vec{X}$,
we have
\beq\lb{sigma,C-prim}     %eq.33
\sig^\pr\equiv \sig(\vec{X}^\pr) = \Lam\sig\Lam^T,\qquad
C^\pr \equiv C(\vec{X}^\pr) = \Lam C\Lam^T.
\eeq
It is now seen that if the transformation is non-degenerate, det$\Lam \neq
0$, then  the equality in the relation (\ref{SUR}) remains invariant, i.e.
$\det\sig= \det C\,\,\longrightarrow \,\, \det\sig^\pr = \det C^\pr$.  This
implies that in the canonical case of $X_1=p$, $X_2=q$ the equality in
(\ref{SUR}) is invariant under linear canonical transformations. The
equality in the Heisenberg relation is not invariant under linear
transformations.

 In the Heisenberg and the Schr\"odinger inequalities the second moments of
{\it two} observables $X_{1,2}$ are involved.  However two operators never
close an algebra [An exception is the Heisenberg--Weyl algebra $h_1$ due to
the fact that the third operator closing the algebra is the identity
operator: the equality in the $p$-$q$ Schr\"odinger relation (but not in the
Heisenberg one) is invariant under the linear canonical transformations].
Therefore the equality in these uncertainty relations is not invariant under
the general transformations in the algebra to which $X_{1,2}$ may belong.
For $n$ generators of Lie algebras it is desirable to have uncertainty
relations invariant under algebra automorphisms, in particular under the
corresponding Lie group action in the algebra.

Such invariant uncertainty relations turned out to be those of Robertson
\ci{Rob} and of Trifonov and Donev \ci{TriDon}. The Robertson relation for
$n$ observables $X_1,X_2,\ld X_n$ reads  ($i,j,k = 1,2,\ld n$)
\beq\lb{RUR}     %eq.34
{\rm det}\,\sigma(\vec{X})\,\, \geq \,\,{\rm det}\, C(\vec{X}),
\eeq
where $\sigma_{ij} = \Dlt X_i X_j$, and $C_{kj} = -i\la[X_k,X_j]\ra/2$.
With minor changes the Robertson proof of (\ref{RUR}) is provided in the
Appendix. The minimization of (\ref{RUR}) is considered in detail in
\ci{Trif97}, the minimizing states being called Robertson intelligent
states or Robertson {\it optimal US}. A pure state minimize (\ref{RUR}) if
it is an eigenstate of a real combination of the observables. For odd $n$
this is also a necessary condition. Robertson optimal US 
exist for a broad class of observables, the simplest example being given
by the well known $N$-modes Glauber CS $|\vec{\alf}\ra = |\alf_1\ra\,
|\alf_2\ra,\ld |\alf_N\ra$,  and by the $N$-modes  canonical SS
$|\vec{\alf},u,v\ra$ (constructed in \ci{MMT,HMMT} with no reference to 
the Robertson relation). A more general examle is given by the group-related
CS $\{T(g),\Psi_0\}$ when $|\Psi_0\ra$ is eigenstate of a (real) Lie algebra
element \ci{Trif97}. If in addition $|\Psi_0\ra$ is the lowest (highest)
weight vector (the case of semisimple Lie groups \ci{ZhaFenGil}) then these
CS minimize (\ref{RUR}) for the Hermitian components of Weyl generators as
well \ci{Trif97}. On the example of the  $SU(2)$ and $SU(1,1)$ CS, eqs.
(\ref{SU2CS}) and (\ref{SU11CS}),  the above minimization properties can be
checked by direct calculations. In the case of one-mode and two-mode boson
representations of $su(1,1)$ the above properties mean that squeezed Fock
states minimize (\ref{RUR}) for the three generators $K_i$, but  
squeezed vacuum in addition minimizes (\ref{SUR}) for $K_1$ and $K_2$. 

The number of the Hermitian components of Weyl generators (of a semisimple
Lie group) is even.  For the even number $n$ of observables the Robertson
inequality (\ref{RUR}) is minimized in a state $|\Psi\ra$ if the latter is an
eigenstate of $n/2$ complex linear combinations of $X_j$.  For these
minimizing states the second moments of $X_i,\,X_j$ can be expressed in
terms of the first moments of their commutators.  In that purpose and
keeping the analogy to the case of canonical SS (\ref{2a}) we define
$\tilde{a}_\mu = X_\mu +iX_{\mu+N}$ and write down the $n/2\equiv N$ complex
combinations as ($\mu,\nu = 1,2,\ld, N$)
\beq\lb{}     %eq.35
A_\mu(u,v) := u_{\mu\nu}\tilde{a}_\nu +
v_{\mu\nu}\tilde{a}^\dg_\nu = \beta_{\mu j}X_j ,
\eeq
where $\beta_{\mu\nu} = u_{\mu\nu}+v_{\mu\nu}$,\, $\beta_{\mu,s+\nu} =
i(u_{\mu\nu} - v_{\mu\nu})$.
Then after some algebra we get that in the eigenstates $|\vec{z},u,v\ra$ of
$A_\mu(\beta)$ the following general formula holds,
\bear\lb{nOUSmeans}     %eq.36
\sig(\vec{X};z,u,v) = {\cal B}^{-1}\left( \bt{cc} $0$ & $ \td{C} $\\
$\td{C}^{\rm T}$& $0$ \et \right)
{\cal B}^{-1}{}^{\rm T},\\[2mm]
\td{C}_{\mu\nu} = \frac 12\la[A_\mu,A_\nu^\dg]\ra,\quad 
{\cal B} =  \left(\bt{cc} $u+v$&$i(u-v)$\\
$u^* + v^*$&$i(v^* - u^*)$\et\right). \nn
\eear
Note that $u,\,v$ and $\td{C}$ are $N\times N$ matrices, $\beta$ is an
$N\times n$ matrix, while ${\cal B}$ is $n\times n$.  We suppose that
${\cal B}$ is not singular. For two observables, $n=2$, we have
$\beta_{11} = u+v$, $\beta_{12}= i(u-v)$ and formula (\ref{nOUSmeans})
recovers (\ref{OUSmeans}).

The Robertson inequality relates the determinants of two $n\times n$
matrices $\sig$ and $C$. These are the highest order {\it characteristic
coefficients} of the two matrices \ci{Gantmaher} which are invariant under
similarity transformations of the matrices. Then from (\ref{sigma,C-prim})
we see that $\det\sig$ and $\det C$ are invariant under the orthogonal
transformations of the observables. However, one can see, again from the
transformation law (\ref{sigma,C-prim}), that the equality in (\ref{RUR}) is
invariant under {\it any}   nondegenerate linear transformations of the
$n$ observables.  Now we recall \ci{Gantmaher} that for an $n\times n$ matrix
$M$ there are $n$ invariant characteristic coefficients $C_r^{(n)}$, $r=
1,2,\ld, n$, defined by means of the secular equation 
\beq\lb{chareqn}     %eq.37
0 = \det(M - \lam) = \sum_{r=0}^{n} C^{(n)}_r(M)(-\lam)^{n-r}.
\eeq

The characteristic coefficients $C^{(n)}_r$ are equal to the
sum of all principle minors ${\cal M}(i_1,\ldots,i_r;M)$ of order $r$.  One
has $C^{(n)}_0 = 1$, $C^{(n)}_1 = {\rm Tr}\,M = \sum m_{ii}$ and $C^{(n)}_n
= \det M$. For $n=3$ we have, for example, three principle minors of order
$2$.  In these notations Robertson inequality (\ref{RUR}) reads
$C^{(n)}_n\left(\sig(\vec{X})\right) \geq C^{(n)}_n\left(C(\vec{X})\right)$.
It is important to note now two points: (1) the uncertainty matrix
$\sig(\vec{X})$ and the mean commutator matrix $C(\vec{X})$ are nonnegative
definite and such are all their principle minors; (2) The principle minors
of $\sig(\vec{X})$ and $C(\vec{X})$ of order $r$ can be regarded as
uncertainty matrix and mean commutator matrix for $r$ observables $X_{i_1},
\ld, X_{i_r}$ correspondingly. Then all characteristic coefficients of the
two matrices obey the inequalities \ci{TriDon}
\beq\lb{CUR}     %eq.38
C^{(n)}_r\left(\sig(\vec{X})\right) \,\geq\,
C^{(n)}_r\left(C(\vec{X})\right),\quad r=1,2,\ld,n.
\eeq
These invariant relations can be called {\it characteristic uncertainty
relations}. The Robertson relation (\ref{RUR}) is one of them and can be
called the $n^{\rm th}$-order characteristic inequality.

The minimization of the {\it first order} inequality in (\ref{CUR}),
Tr$\,\sig(\vec{X}) = {\rm Tr}\,C(\vec{X})$,  can occur in the case of
commuting operators only since Tr\,$C(\vec{X})\equiv 0$. Important examples 
of minimization of the {\it second order} inequality were pointed out in
\ci{TriDon} -- the spin and quasi spin CS $|\tau;j\ra$ and $|\xi;k\ra$
minimize the second order characteristic inequality for the three
generators $J_{1,2,3}$ and $K_{1,2,3}$ correspondingly. We have already
noted that these group-related CS minimize the {\it third order}
inequalities too, so their characteristic minimization "ability" is
maximal. 
The analysis of the solutions of the eigenvalue equation $[uK_- + vK_+ +
wK_3]\,|\Psi\ra = z|\Psi\ra$ shows (see Appendix) that the CS $|\xi;k\ra$
are {\it the unique states} which minimize simultaneously the second and the
third order characteristic inequalities for $K_{1,2,3}$ and there are no
states which minimize the second order inequality only. Thus the
minimization of the characteristic inequalities (\ref{CUR}) of order $r < n$
can be used for {\it finer classification} of group-related CS with
symmetry. It turned out (see the Appendix) that the uniqueness of these 
states follows also 
from the requirement to minimize simultaneously (\ref{RUR}) for the three
generators and (\ref{SUR}) for the Hermitian components of $K_-$. \\[0mm]

All the above characteristic inequalities \footnote{Let us note that other
types of uncertainty relations, e.g. the entropic and the
parameter-based ones, are also considered in the literature \cite{drugi}.}
relate combinations 
$C_r^{(n)}(\sig(\vec{X};\rho))$ of second moments of $X_1,\ld,X_n$ in a
(generally mixed) state $\rho$ to the combinations
$C_r^{(n)}(C(\vec{X};\rho))$ of first moments of their commutators in the
same state. It turned out that these relations can be extended to the case
of {\it several state} in the following way. From the derivation of the
characteristic inequalities (\ref{CUR}) (see Appendix) one can deduce that
they are valid for any nonnegative definite matrix ${\cal S}+i{\cal C}$ with
${\cal S}$ nonnegative definite and symmetric and ${\cal C}$ --
antisymmetric.  Well, the finite sum $\sum_md_m\sig_m$, $d_m\geq 0$, of
nonnegative and symmetric matrices is nonnegative and symmetric, and the
finite sum of antisymmetric matrices is again antisymmetric. And if 
$\sig_m+iC_m\geq 0$ their finite sum is also nonnegative. Thus we obtain the
{\it extended} characteristic uncertainty inequalities 
\beq\lb{extendCUR}     %eq.39
C_r^{(n)}\left({\textstyle\sum_m}d_m\sig_m\right) \,\geq\,
C_r^{(n)}\left({\textstyle\sum_m}d_m C_m\right),
\eeq
where $d_m$ are arbitrary real nonnegative parameters.
 Here $\sig_m$ and $C_m$, $m=1,2,\ld$, may be the uncertainty and the mean
commutator matrices for $\vec{X}$ in states $\rho_m$ or the uncertainty
and the mean commutator matrices of different sets of $n$ observables
$\vec{X}^{(m)}$ in the same state $\rho$. For $r=n$ in (\ref{extendCUR}) we
have the extension of the Robertson relation to the case of several states
and/or several sets of $n$ observables. In the first case the
extension reads
\beq\lb{extendRUR}     %eq.40
{\rm det}\left({\textstyle\sum_m}d_m\sig(\vec{X},\rho_m)\right)\, \geq\,
{\rm det}\left({\textstyle\sum_m}d_mC(\vec{X},\rho_m)\right).
\eeq
Since $\det\sum \sig_m\neq \sum\det\sig_m$ these are indeed {\it new}
uncertainty inequalities, which extend the Robertson one to several states.
We note that the extended relations (\ref{extendCUR}), (\ref{extendRUR}) are
{\it invariant} under the nondegenerate linear transformations of the
operators $X_1,\ld,X_n$. If the latter span a Lie algebra then we obtain the
invariance of (\ref{extendCUR}) under the Lie group action in the algebra.
If for several states $|\psi_m\ra$, $m=1,2,\ld$, the inequality
(\ref{extendRUR}) is minimized, then it is minimized also for the 
group-related CS $U(g)|\psi_m\ra$ as well, $U(g)$ being the unitary rep of
the group $G$. In the simplest case of two observables $X,\,Y$ and two
states $|\psi_{1,2}\ra$ which minimize Schr\"odinger inequality (\ref{SUR})
eq.  (\ref{extendRUR}) produces
\bear\lb{extendSUR}     %eq.41
\frac 12 \left[\sig_{XX}(\psi_1)\sig_{YY}(\psi_2) +
\sig_{XX}(\psi_2)\sig_{YY}(\psi_1)\right] - \sig_{XY}(\psi_1)\sig_{XY}(\psi_2)
\nn \\
\geq -\frac 14\la\psi_1|[X,Y]|\psi_1\ra \la\psi_2|[X,Y]|\psi_2\ra,
\eear
where, for convenience, $\sig_{XX}(\psi)$ denotes the variance of $X$ in
$|\psi\ra$ and $\sig_{XY}(\psi)$ denotes the covariance. The more
detailed analysis (to be presented elsewhere) shows that this
uncertainty relation holds for every two states.  For $\psi_1=\psi_2$
the {\it new inequality} (\ref{extendSUR}) recovers that of Shcr\"odinger.
One can easily verify (\ref{extendSUR}) for $p$ and $q$ and any two Fock
states $|n\ra$ and/or Glauber CS $|\alf\ra$ for example. The relation is
minimized in two squeezed states $|\alf_1,u,v\ra$ and
$|\alf_2,u,v\ra$, Im$(uv^*)=0$. Looking at (\ref{extendSUR}) and (\ref{SUR})
one feels that, to complete the symmetry between states and observables, the
third inequality is needed (for one observable and two states), namely
\beq\lb{extendSUR2}     %eq.42
\sig_{XX}(\psi_1)\sig_{XX}(\psi_2) \geq
\left|\la\psi_2|X^2|\psi_1\ra\right|^2-\sig_{XX}(\psi_1)
\la\psi_2|X|\psi_2\ra^2 - \sig_{XX}(\psi_2)\la\psi_1|X|\psi_1\ra^2. 
\eeq
Relations (\ref{SUR}) and (\ref{extendSUR2}) both follow from the Schwarz
inequality, while (\ref{extendSUR}) is different.\\

It is worth noting that every extended characteristic inequality
can be written down in terms of two new positive quantities the sum of
which is not greater than unity. Indeed, let us put
\beq\lb{P_r}     %eq.43
C_r^{(n)}(\sig(\vec{X},\rho)) = \alf_r (1-P_r^2),
\eeq
where $0\leq P_r^2 \leq 1$ (i.e. $1-P_r^2 \leq 1$) and $\alf_r \neq
0$. For $r=n$ eq. (\ref{P_r}) reads (omitting index $r=n$)
det$\sig(\vec{X},\rho) = \alf\,(1-P^2)$. $\alf_r$ may be viewed as scaling
parameters. Then we can put $C_r^{(n)}(C(\vec{X},\rho)) = \alf_r V_r^2$ and
obtain from (\ref{CUR}) the inequality for $P_r$ and $V_r$

\beq\lb{CUR2}     %eq.44
P_r^2(\vec{X},\rho) + V_r^2(\vec{X},\rho) \leq 1,\quad r=1,\ld,n.
\eeq
The equality in (\ref{CUR2}) corresponds to the equality in (\ref{CUR}) (or
(\ref{extendCUR})). For every set of observables $X_1,\ld, X_n$ the
nonnegative quantities $P_r,\,V_r$ are functionals of the state $\rho$ (or
of $\rho_1,\rho_2,\ld$ in the case of extended inequalities
(\ref{extendCUR})). These can be called {\it complementary quantities} and
the form (\ref{CUR2}) of the extended characteristic relations -- {\it
complementary form}. Let us note that $P_r$ and $V_r$ are not uniquely
determined by the characteristic coefficients of $\sig$ and $C$. They depend
on the choice of the scaling parameter $\alf_r$. In the case of bounded
operators $X_i$ (say spin components) the characteristic coefficients of
$\sig$ and $C$ are also bounded.  In that case $\alf_r$ can be taken as the
inverse maximal value of $C_r^{(n)} (\sig)$. In the very simple case of one
state and two operators with only two eigenvalues each the complementary
characteristic inequality (\ref{CUR2}) was recently considered in the
important paper by Bjork et al \ci{Bjork}. In this particular case the
meaning of the complementary quantities $P$ and $V$ was elucidated to be
that of the {\it predictability} ($P$) and the {\it visibility} ($V$) in the
{\it welcher weg} experiment \ci{Bjork}.

Finally we note that as functionals of the states  $\rho$ the characteristic
coefficients of positive definite uncertainty matrix $\sig(\vec{X})$ (then
the coefficients $C_r(\sig(\vec{X},\rho))$ are all positive), can be used
for the construction of {\it distances} between quantum states.  One
possible series of such (Euclidean type) distances
$D^2_r[\rho_1,\rho_2;\vec{X}]$ is \ci{TriDon2}
\bear\lb{dist}     %eq.45    
D^2_r[\rho_1,\rho_2] = C_r(\sig(\vec{X},\rho_1)) +
C_r(\sig(\vec{X},\rho_2)) - 2\left(C_r(\sig(\vec{X},\rho_1))
C_r(\sig(\vec{X},\rho_2))\right)^{\frac 12}\nn\\
\times \, g(\rho_1,\rho_2),\quad
\eear
where $g(\rho_1,\rho_2)$ is any nonnegative functional of $\rho_1,\rho_2$,
such that\, $0\leq g(\rho_1,\rho_2)\leq 1$\,\, and \,$\,\rho_1 = \rho_2
\Leftrightarrow g=1$.\, A known simple such functional ($g$-type
functional) is 
$g(\rho_1,\rho_2) = 
{\rm Tr}(\rho_1\rho_2)/\sqrt{{\rm Tr}(\rho_1^2) {\rm Tr}(\rho_2^2)}$. 
By means of (\ref{extendSUR2}) with any observable $X$ such that $X|\psi\ra
\neq 0$ (continuous or strictly positive $X$, for example) we can construct a
new $g$-type functional  
\beq\lb{g[]}     %eq.46
g(\psi_1,\psi_2;X) =   \frac{\left|\la\psi_2|X^2|\psi_1\ra\right|}
{\sqrt{\la\psi_1|X^2|\psi_1\ra\la\psi_2|X^2|\psi_2\ra}},
\eeq
which can be used for distance constructions, the simplest distance 
being $D^2 = 2\left(1 - g(\psi_1,\psi_2;X)\right)$.
Several other $g$-type functionals are also possible \ci{TriDon2}.
The uncertainty matrix $\sig(\vec{X})$ is positive for examples in the case
of $X_i$ being the quadratures component of $N$ $q$-deformed boson annihilation
operators $a_{q,\mu}$ with positive $q$ \ci{Trif97}.
\vs{5mm}

\section{Conclusion}
  We have briefly reviewed and compared the three ways of generalization of
canonical coherent states (CS) with the emphasis laid on the uncertainty (the
third) way. The Robertson inequality and the other characteristic relations
for several operators \ci{TriDon} are those uncertainty inequalities which
bring together the three ways of generalization on the level of many
observables. The equalities in these relations for the group generators are
invariant under the group action in the Lie algebra. From the Robertson
inequality minimization conditions \ci{Trif97} it follows that all
group-related CS whose reference vector is eigenstate of an element of the
corresponding Lie algebra do minimize the Robertson relation (\ref{RUR}).
The minimization of the other characteristic inequalities (\ref{CUR}) can be
used for {\it finer classification} of group-related CS with symmetry. Along
these lines we have shown that $SU(1,1)$ CS with lowest weight reference
vector $|k,k\ra$ are the unique states which minimize the second order
characteristic inequality for the three $SU(1,1)$ generators. Also, these
are the unique states to minimize simultaneously the Robertson inequality
for the three generators and the Schr\"odinger one for the Hermitian
components of the ladder operator $K_-$.  These statements are valid for 
the $SU(2)$ CS with the lowest (highest) reference vector $|j,\mp j\ra$ as
well. They can be extended to the case of semisimple Lie groups. 

In all so far considered characteristic uncertainty inequalities (the
Schr\"odinger and Robertson relations are characteristic ones) two or more
observables and {\it one state} are involved. It turned out that these
relations, for any $n$ observables, are extendable to the case of two or
more states. We also have shown that the (extended) characteristic
inequalities can be written down in the {\it complementary form} in terms of
two positive quantities less than unity. In the case of two observables with
two eigenvalues each these {\it complementary quantities} were recently
proved \ci{Bjork} to have the meaning of the predictability and visibility
in the welcher weg experiment. The notion of "characteristic complementary
quantities" might be useful in treating complicated quantum systems. It
was also noted that the characteristic coefficients of positive definite
uncertainty matrices can be used for the construction of distances between
quantum states.

\section*{Appendix}

\subsection*{ Robertson Proof of the Relation
               $\det \sig \geq \det C$}

Since the derivation of the characteristic (\ref{CUR}) and the extended
characteristic uncertainty inequalities (\ref{extendCUR}) is based on the
Robertson relation (\ref{RUR}) here we provide the proof of (\ref{RUR})
following Robertson' paper \ci{Rob} with some modern notations.  Let
$X_1,X_2,\ld,X_n$ be Hermitian operators, and $|\psi\ra$ be a pure state.
Consider the squared norm of the composite state $|\psi^\pr\ra = \sum_j \alf_j
(X_j - \la X_j\ra)|\psi\ra$, where $\alf_j$ are arbitrary complex
parameters. One has
\beq\lb{|psi|}     %eq.47
\la\psi^\pr|\psi^\pr\ra = \sum_{jk} \alf^*_k\alf_j
\la\psi|(X_k-\la X_k\ra)(X_j-\la X_j\ra)|\psi\ra
= \sum_{k,j}\alf_k^*S_{kj}\alf_j \equiv {\cal S}(\vec{\alf}^*,\vec{\alf}),
\eeq
where the matrix elements $S_{kj}$ are
$S_{kj} = \la\psi|(X_k-\la X_k\ra)(X_j-\la X_j\ra)|\psi\ra =
\sig_{jk} + i C_{jk}$.
We see that $S =\sig + i C$, where $\sig$ and $C$ are the uncertainty and
the mean commutator matrices  of the operators $X_1,\ld,X_n$ in the state
$|\psi\ra$ (see eq. (\ref{sigma,C})). In Hilbert space we have
$\la\psi^\pr|\psi^\pr\ra = 0$ iff $|\psi^\pr\ra = \sum_j \alf_j(X_j - \la
X_j\ra) |\psi\ra =0$, which means that $|\psi\ra$ is an eigenstate of the
complex combination of $X_j$. Thus the form ${\cal S}$ is nonnegative
definite, which means that the $n\times n$ matrix $S=\sig +iC$ is
nonnegative: all its principle minors are nonnegative \ci{Gantmaher}, in
particular $\det S > 0$. For the case of two operators, $n=2$, one can
easily verify that
\beq\lb{detS}     %eq.48
0\leq\det S =\det(\sig +iC) = \det\sig -\det C,\quad (n=2\,\,\,{\rm only}).
\eeq
This proves the Robertson relation for two observables which was also
derived by Schr\"odinger \ci{SchRob} using the Schwarz inequality.
The property (\ref{detS}) is due to the symmetricity of $\sig$ and
antisymmetricity of $C$ and is valid for $n=2$ only.

For odd $n$, $n\geq 1$, the Robertson inequality $\det\sig \geq \det C$
is trivial, since the determinant of an antisymmetric matrix of odd
dimension vanishes identically. For even $n=2N$ and $n>2$ we follow the
proof of Robertson \ci{Rob}, using however some notions from the present
matrix theory \ci{Gantmaher}. One considers the regular sheaf (bundle)
of the matrices $\sig$ and $\eta = iC$, $\eta - \lam\sig$, supposing
$\sig > 0$. There exist congruent transformation (by means of the so called
sheaf principle matrix $Z$, $\det Z\neq 0$), which brings both
matrices to the diagonal form -- $\sig$ to the unit matrix, $\sig^\pr =
Z^T\sig T = 1$ and $\eta^\pr = {\rm diag}\{\lam_1,\lam_2,\ld,\lam_{2N}\}$,
where $\lam_i$ are the $2N$ roots of the secular equation
$\det(\eta-\lam\sig) = 0$. The product of all roots equals
$\det\eta/\det\sig$. From $\det (\eta-\lam\sig) = \det(\eta-\lam\sig)^T =
\det (\eta+\lam\sig)$ (since $\eta^T=-\eta$ and $n=2N$) it follows that the
polynomial $\det(\eta-\lam\sig)$ contains only even powers of $\lam$,
$\det(\eta-\lam\sig) = \det\eta + \ld + (-\lam)^{2N}\det\sig = 0.$ This
means  that the $2N$ real roots $\lam_j$ are equal and opposite in pairs.
Denoting positive routs as $\lam_\mu$, $\mu = 1,\ld,N$ and negative roots as
$\lam_{\mu+N} = -\lam_\mu$ one writes\\[-3mm]
\beq\lb{detC}     %eq.49
\det\eta = (-1)^N\det C = (-1)^N\prod_\mu\lam_\mu^2\,\det\sig.
\eeq\\[-2mm]
On the other hand the Hermitian matrix $\sig + \eta = \sig +iC$ is positive
definite and after the diagonalization takes the form 
\beq\lb{s^pr+eta^pr}     %eq.50
\sig^\pr +\eta^\pr = {\rm diag}\{1+\lam_1,\ld,
1+\lam_{2N}\} = {\rm diag}\{1+\lam_1,1-\lam_1,\ldots,1+\lam_N,1-\lam_N\}.
\eeq
%\beq\lb{detS3}
%\det(\sig+\eta) = (\det Z)^{-2}\prod_\mu^N (1-\lam_\mu^2) > 0,
%\eeq
The diagonal matrix $\sig^\pr +\eta^\pr$ is again nonnegative definite,
i.e. all the elements on the diagonal are nonnegative, which implies that
$\lam^2_\mu \leq 1$, $\mu = 1,\ld, N$. Then eq.  (\ref{detC}) yields the
Robertson inequality $\det \sig \geq \det C$. End of the proof. \\[0mm]

{\bf Remarks:} 
(a) Robertson considered the case of pure states only.  However one can
see from the proof that his relation holds for mixed states as well; 
(b) It is seen from the above proof that the inequality $\det
\sig \geq \det C$ holds for any two real matrices $C$ and $\sig$, one of
which is antisymmetric ($C$), the other -- symmetric and nonnegative
definite and such that Hermitian matrix $\sig + iC$ is again nonnegative;
(c) If the matrices $\sig_j$ and $C_j$, $j=1,2,\ld,m$, obey the requirements
of (b) then $\det(\sig_1+\sig_2+\ld) \geq \det(C_1+C_2+\ld)$ since (as one
can easily prove) the sum of nonnegative $\sig_j + iC_j$ is again a
nonnegative matrix.  These observations have been used in establishing the
extended characteristic relations (\ref{extendCUR}) for several
states and in formulating the remark (a) as well.
\vs{3mm}

\subsection*{The $SU(1,1)$ CS $|\xi;k\ra$ are the Unique States
   Which Minimize the Characteristic Inequalities for the Three Generators}

For the three generators $K_i$ of $SU(1,1)$ there are two nontrivial
characteristic uncertainty inequalities corresponding to $r=n=3$ and
$r=n-1=2$ in (\ref{CUR}). The third order characteristic UR is minimized in
a pure state $|\psi\ra$ iff $|\psi\ra$ is an eigenstate of a real
combination of $K_i$, i.e. iff $|\psi\ra = |z,u,v,w;k\ra$ obey the equation
\beq\lb{CUS3}     %eq.51
[uK_- + vK_+ + wK_3]\,|z,u,v,w;k\ra = z|z,u,v,w;k\ra
\eeq
with real $w$ and $v=u^*$. The second order characteristic UR is
minimized iff $|\psi\ra$ is an eigenstate of complex combinations
of all three pair $K_i$-$K_j$ simultaneously, i.e. iff
\beq\lb{CUS2}     %eq.52
\left.\begin{tabular}{l}
$[u_1K_- + v_1K_+ +w_1K_3]\,|\psi\ra = z_1|\psi\ra,\quad w_1=0,$\\
$\,[u_2K_- + v_2K_+ +w_2K_3]\,|\psi\ra = z_2|\psi\ra,\quad v_2=u_2,\,
w_2\neq0,$\\ $\,[u_3K_- + v_3K_+ +w_3K_3]\,|\psi\ra = z_3|\psi\ra,\quad
v_3=-u_3,\, w_3\neq0,$
\end{tabular}\right\}
\eeq
where the complex parameters
$u_1,\,v_1,\,u_2,\,w_2,\,u_3,$ and $w_3$ shouldn't vanish and
$z_1,\,z_2,\,z_3$ may be arbitrary. 
To solve this system it is convenient to use BG analytic rep (\ref{BGrep}).
Let us start with the first equation in (\ref{CUS2}). Its normalizable
solutions $|z_1,u_1,v_1;k\ra$ for $k=1/2,1,\ld$ were found in \ci{Trif94}.
They are normalizable for $|u_1|>|v_1|$ only and in BG rep have the form (up
to the normalization constant)
\beq\lb{K1-K2CUS}     %eq.53
 \Phi_{z_1}(\eta;u_1,v_1) = e^{-\eta\sqrt{-v_1/u_1}}\,_1F_1\left(k+\frac{z_1/2}
{\sqrt{-u_1v_1}};2k;2\eta\sqrt{-v_1/u_1}\right),
 \eeq
where (for any $|u_1|>|v_1|$) the eigenvalue $z_1$ is arbitrary complex
number. Here the complex 
variable in the BG rep (\ref{BGrep}) is denoted by $\eta$. For $k<1/2$ a
second normalizable solution exist of the form
\beq\lb{secondsol}     %eq.54
 \Phi^\pr_{z_1}(\eta;u_1,v_1) = \eta^{1-2k}\,e^{-\eta\sqrt{-v_1/u_1}}\,
 _1F_1\left(\frac{z_1/2} {\sqrt{-u_1v_1}}-k+1;2(1-k);2\eta\sqrt{-v_1/u_1}
 \right),
\eeq
In order to obtain second order $SU(1,1)$ characteristic US we have to subject 
the solution (\ref{K1-K2CUS}) to obey the rest two equations in (\ref{CUS2}).
Let us try to obey the second one.
Since $u_1\neq 0$ we can write $K_-|z_1,u_1,v_1;k\ra = (z_1-v_1K_+)
|z_1,u_1,v_1;k\ra/u_1$ and substitute into the second equation to obtain
\beq\lb{K_3|zuv>}     %eq.55
K_3|z_1,u_1,v_1;k\ra = \frac{1}{w_2}[z_2-\frac{u_2}{u_1}z_1
+(v_1\frac{u_2}{u_1}-u_2)K_+]|z_1,u_1,v_1;k\ra.
\eeq
In BG rep (\ref{BGrep}) this is a first order equation which the function
(\ref{K1-K2CUS}) has to obey. By equating the coefficients of the terms
proportional to $\eta^n$, $n=0,1,\ld$, we obtain after some manipulations the
necessary conditions (a) $k + z_1/2\sqrt{-u_1v_1} = 0$; (b) $k = z_2/w^2 -
u_2z_1/u_1w_2$ and (c) $u_2(1-v_1/u_1) = w_2\sqrt{-v_1/u_1}/2$. 
The first condition requires the relation between the parameters
$z_1,\,u_1,\,v_1$ and reduces the "wave function" (\ref{K1-K2CUS}) to
\beq\lb{CSunique}     %eq.56
\Phi_{z_1}(\eta;u_1,v_1) = \exp\left[-\eta\sqrt{-v_1/u_1}\right],
\eeq 
which is just the CS $||\xi;k\ra$ in BG rep with $\xi = -\sqrt{-v_1/u_1}$.
The second condition is always satisfied by $z_2 = kw_2 + u_2z_1/u_1$,
$u_2,\,w_2$ remaining arbitrary.  Thus {\it it is the CS $|\xi;k\ra$
only}, $k=1/2,1,\ld$, which minimize simultaneously the Schr\"odinger
inequality for $K_1,\,K_2$ and $K_1,\,K_3$.

Next it is a simple (but not short) exercise to check that
$\exp\left[-\eta\sqrt{-v_1/u_1}\right]$ satisfy the third equation in
(\ref{CUS2}) with
$w_3 = w_2(z_3u_1-iu_3z_1)/(u_1z_2-u_2z_1)$,
$z_3 = i(u_3/u_2u_1)\,(u_1+v_1z_1)/(u_1z_2-u_2z_1) + iu_3z_1/u_1$,
 ($u_2,z_2,\,u_3$ being free) and the eigenvalue eq. (\ref{CUS3}) with
$v=u^*$ and real $w$, $w = (-uv_1 + u^*u_1)/\sqrt{-u_1v_1} = w(u_1,v_,u)$.
One can see that for every given $\xi=-\sqrt{-v_1/u_1}$ the equation
Im$[w(u_1,v_1,u)] = 0$ can be solved with respect to $u$, the solution being
not unique: $u = |u|\exp\left[\pi/4 - {\rm arg}\xi/2\right]$, $|u|$ being
arbitrary. So the family of CS $|\xi;k\ra$ is the unique family of states
which minimize the third and the second order characteristic US
simultaneously. If we subject the function (\ref{K1-K2CUS}) directly to 
(\ref{CUS3}) we will get again (\ref{CSunique}).

In the case of $SU(1,1)$ characteristic US for $K_i$ in rep (\ref{1moderep})
($k = 1/4,3/4$) we have to consider the two solutions
(\ref{K1-K2CUS}) and (\ref{secondsol}).  The consideration gives no new
result - again the eqs. (\ref{CUS3}) and (\ref{CUS2}) are satisfied by
$\exp\left[-\eta\sqrt{-v_1/u_1}\right]$ only.

Similar results can be obtained for the minimization of (\ref{RUR}) 
in CS $|\tau;j\ra$ using for example their own analytic representation
and the results of paper \ci{Brif97}.

\subsection*{Acknowledgment}
This work is partially supported by the National Science Fund of the 
Bulgarian Ministry of Education and Science, Grant No F-644/1996.

\vspace{5mm}

{\footnotesize 
Misprints in v. 4 (here, in v. 5, corrected):
 
In eq.(5):  $\frac{2\hbar}{m} \lrar \left(\frac{2\hbar}{m\ome_0}\right)^{1/2}$,

In eq.(36): $[A_\mu,A_\nu] \lrar [A_\mu,A_\nu^\dg]$.}

\end{document}